\documentclass[12pt,preprint]{aastex}

\usepackage{emulateapj5}
\usepackage{psfig}

\slugcomment{accepted for publication in the Astrophysical Journal}

\shorttitle{The mass donor of Sco X-1}
\shortauthors{Steeghs \& Casares}

\begin{document}


\title{The mass donor of Scorpius X-1 revealed}


\author{D.Steeghs}
\affil{Department of Physics \& Astronomy, University of Southampton, Highfield, Southampton SO17 1BJ, UK}
\email{ds@astro.soton.ac.uk}
\and
\author{J.Casares}
\affil{Instituto de Astrofisica de Canarias, Via Lactea, La Laguna, E-38200, Santa Cruz de Tenerife, Spain}
\email{jcv@ll.iac.es}

\begin{abstract}

We present the first detection of the mass donor star in the prototype
X-ray binary  Scorpius   X-1.   Phase resolved  spectroscopy  revealed
narrow emission  lines components from  the irradiated secondary star.
Radial velocity fits to the Bowen blend emission are used to establish
an absolute  orbital ephemeris and  derive  an accurate value  for the
systemic velocity; $\gamma= -113.8 \pm 0.6$ km s$^{-1}$.
Emission  from the  irradiated  front side of   the secondary leads to
solid limits to the radial velocity of the mass donor of $87 \la K_2
\la 148$ km   s$^{-1}$. In conjunction  with  an  upper limit  to  the
velocity of the primary  $K_1 \le 53$  km s$^{-1}$,  we derive a  firm
limit on the  mass ratio of  Sco X-1 of $q \la  0.61$. 
A likely  set of system parameters  satisfying the various constraints
as well  as the recent inclination estimate  by Fomalont et al. (2001)
is $K_1=40$ km s$^{-1}$, $K_2=133$ km s$^{-1}$, $q=0.30$ corresponding
to component masses of $M_1=1.4M_{\sun}$  , $M_2=0.42 M_{\sun}$ for an
orbital inclination of $38^{\circ}$.

\end{abstract}



\section{Introduction}

Scorpius X-1 is the prototype low-mass X-ray binary (LMXB), consisting of a
neutron  star accreting matter   from  a Roche lobe  filling companion
star.  It is the brightest persistent X-ray source  in the sky and has
been the  target of detailed  analysis  for the last  35 years  in all
energy bands,  from radio to X-rays. An  orbital period of 18.9 hr was
reported  by Gottlieb,  Wright \&  Liller  (1975)  from brightness eye
estimates on  1068 plates during  the interval 1890-1974.  The optical
B-band light curve shows a 0.13 mag amplitude modulation (Augusteijn et
al.  1998) which is being interpreted  as the  visibility of the X-ray
heated companion.  Radial velocity curves of HeII $\lambda$4686 and HI
confirmed the    orbital  period  and   indicated  that  the  inferior
conjunction of the emission line  regions are close to the photometric
minimum and, hence, they    must  originate near the compact    object
(e.g. LaSala \& Thorstensen 1985). 
Bradshaw et al. (1999) measured the trigonometric  parallax of Sco X-1
using VLBA radio observations  and deduce a   distance of $2.8  \pm 0.3$
kpc.   Intensive radio  observations  were also   used  to measure the
properties of  the  twin radio  lobes   around the source  which   are
inclined at an  angle of $44  \pm 6^{\circ}$  relative to  the line of
sight (Fomalont et al. 2001).
Despite intensive  efforts, spectroscopic signatures  of the companion
star  have not yet been observed  in optical-IR  data (e.g. Schachter,
Filippenko  \& Kahn  1989; Bandyopadhyay et   al. 1997,1999). Here  we
present  high  resolution spectroscopy of Sco   X-1 which enable us to
detect the companion star for the first time.

\section{Observations and reduction}

The phase resolved spectroscopy of Scorpius X-1  was obtained with the
4.2m William  Hershel Telescope on   La Palma in  combination with the
ISIS double beam spectrograph. The R1200B grating centred at 4601\AA~
delivered  a wavelength coverage  of  4150-5050\AA~at 0.45\AA/pixel on
the 2048x4096 pixel EEV CCD detector. A 1.2'' slit width resulted in a
spectral resolution of 0.84\AA, corresponding to 55 km s$^{-1}$ at our
central wavelength.  The  data  were part  of a  campaign to  look for
circular polarisation  from   accreting binaries,  and to this   end a
quarter-wave  plate  and  a  Savart calcite slab  were  placed  in the
optical path (see Tinbergen  \& Rutten (1992) for  more details).  The
slit   position  angle was adjusted   to  remain close  to parallactic
throughout the observations.  A total  of 137 spectra were obtained in
the course of three consecutive nights covering  75\% of the 18.9 hour
orbital cycle. A journal of observations can be found in Table 1.


The insertion of  the calcite slab  in the light  path results in  the
projection of two  target beams on the  detector.  The intensity ratio
then  represents the   fraction of  circular   polarisation.  For  the
analysis presented in  this    Letter, we  ignored    the polarisation
characteristics of the light, and combined  the extracted spectra from
both beams.  Frames  were first de-biased   using a median  bias frame
obtained by combining  10-15 individual bias  exposures.  The overscan
strip was then used  to subtract any  residual bias from each frame. A
median of 29 Tungsten exposures was constructed at each night in order
to  perform  flat  field correction.   Each  beam  was then  optimally
extracted. Some frames were  affected  by transient CCD  pickup  noise
producing  variable bias structure at the  level of 3-5 ADU. For those
frames affected (around  8 on the second   night and 10  on the third
night), we   introduced  an  additional  reduction  step   before  optimal
extraction   that  subtracted the pickup     noise using a  2nd  order
polynomial  fit   to each CCD row.    Regular   CuAr arclamp exposures
allowed   us  to  establish  an accurate   wavelength   scale for each
spectrum.  Finally,  a   wide  slit  exposure   of  the flux  standard
BD28+4211  (Oke 1990) was used to   perform flux calibration using the
MOLLY data analysis package.  The two extracted spectra for each frame
were then added to produce the final set of 137 calibrated Sco X-1 spectra.

\section{Analysis}

\subsection{The emission line spectrum}

The optical spectrum  of Sco  X-1   is dominated by  strong and  broad
emission lines from  the Balmer series  as well as HeI/HeII.   However,
closer inspection revealed  a large number  of weaker and very  narrow
emission line features  that showed significant  Doppler  motions as a
function of orbital phase. In particular,  the strong Bowen blend near
4640\AA~was  resolved  into an underlying  broad  component as well as
several narrow  components from   the individual  Bowen   transitions.
Broad  and highly variable  emission around 4640\AA~was already noted
by Sandage et  al.  (1966) in the  paper discussing the identification
of the optical counterpart of Sco X-1.   A more detailed investigation
of the  same data  by Westphal,  Sandage \&  Kristian (1968) did  also
indicate  complex  and strongly time  dependent  fine structure in the
broad emission structures  and some preliminary evidence for  velocity
changes in the strong Balmer and HeII lines.
More recently, Bowen emission  in Sco X-1  was noted by Willis et  al.
(1980)  while Schachter, Filippenko  \& Kahn (1989)  discuss the Bowen
fluorescence  process in  Sco X-1  in  detail.  However, neither  had the
spectral resolution to resolve the  individual components. We will see
in the next sections   that  these narrow   emission features  can  be
attributed to  the irradiated companion star,  thus revealing a direct
signature from  the mass  donor  in Sco X-1 for  the  first  time.  In
Figure 1 we present the average normalised  spectrum of Sco X-1 in the
rest frame  of the companion  star to facilitate the identification of
the  narrow features.  Most  lines were identified using the planetary
nebulae emission line  lists from Kaler (1976)  as well as  the Kurucz
atomic line list.  Many narrow lines from OII are present as well as a
few SiIII, CII and NIII transitions.


\subsection{Bowen emission from the donor}

In  order   to establish  the   origin  of  the  narrow  emission line
components, we measured the velocity of the narrow Bowen components as
a function of phase.  As can be seen in  Figure 1, the strongest Bowen
components are the two NIII components at  4634.13\AA~and 4640.64\AA, and
the CIII components  at 4647.4/4650.1\AA. In  Figure 2, we display  the
Bowen blend  and  HeII4686   emission   in the    form of  a   trailed
spectrogram. All  the individual spectra were  first normalised to the
continuum using a 3rd order polynomial fit. The two NIII components as
well as the CIII component at 4647\AA~clearly move  in phase with each
other, and are   extremely  narrow.  In  the background  more  diffuse
emission can be seen  to move roughly in  anti-phase with these  sharp
components.  For comparison, the broad and   weakly double peaked HeII
line shows completely different line kinematics.


The velocities of the various components  were measured using multiple
Gaussian fits to the individual spectra.  At first we fitted the Bowen
profiles using 3  narrow  Gaussians corresponding  to the 3  strongest
Bowen components as well as an underlying broad component. The FWHM of
the Gaussians were determined  from a fit  to the average  Bowen blend
profile and fixed  to 80 km s$^{-1}$ for  the narrow components and 1550 km s$^{-1}$
for the broad  component. The position and  strength for each Gaussian
was a free parameter for each  individual spectrum. The derived radial
velocities of the   sharp components were  identical within  the error
bars. We thus decided to  reduce the number of  free parameters in our
fits and give the three narrow Gaussians a common radial velocity, but
an independent   strength. The 6 parameters  were  optimised  for each
observed spectrum through least square fitting to determine the common
velocity  of the  three narrow  components  and its associated  formal
error.


Figure  3 plots the   derived  radial  velocity   curve of the   Bowen
components together with a least squares sinusoidal fit.  The best fit
gives a velocity amplitude of  $77.2 \pm 0.4$  km s$^{-1}$ with a mean
velocity of $-113.8 \pm 0.6$ km s$^{-1}$.  Given the extreme narrowness
of the  components, they cannot originate from  the accretion  flow or
hot-spot that is expected  to be  responsible  for the broad  emission
line  components.   Such  narrow   emission components  are,  however,
regularly seen in the  line  profiles of cataclysmic  variables  where
they originate  from the irradiated front  side of the  companion star
(e.g. Harlaftis  et al.  1999, Schwope et  al.  2000).   This has also
been   observed in  the intermediate mass  X-ray binary   Her  X-1 (Still et
al. 1997), where narrow Bowen blend components from the mass donor are
present.  An origin on the mass  donor is furthermore supported by the
fact that the crossing of  the emission components  from blue shift to
red shift    occurs near minimum light, and    that the strength of the
emission is weakest then, since we are  looking at the unirradiated backside
of  the   secondary.      The   inferred    systemic    velocity    of
$\gamma=-113.8\pm0.6$  km   s$^{-1}$ furthermore  agrees   well with  previous
estimates using the strong emission  lines (e.g. Lasala \& Thorstensen
1985).  We thus have established  a radial velocity  curve of the mass
donor  in Sco  X-1 for  the  first  time.  The radial velocity   curve
directly gives us an  unbiased value for the  systemic velocity of the
binary and  establishes the correct absolute  orbital phases. From the
sinusoidal fit we established the following orbital ephemeris which we
will use throughout this paper;
\[
T_0 (HJD) = 2451358.568(3) + 0.787313(1) E
\]
\noindent where we took the value for the orbital period as derived by Gottlieb, Wright \& Liller (1975).
Using our  ephemeris, the phase  of  minimum light  as established  by
Gottlieb, Wright  \&  Liller  corresponds  to  an  absolute  phase  of
$\phi=0.04 \pm  0.04$, where we have allowed  for the uncertainties in
both $T_0$ and $P$.  This confirms  that the photometric modulation is
indeed related to the X-ray heated hemisphere of the donor star. $T_0$
values   based on the inferior  conjunction  of the HeII $\lambda$4686
emission, on the other hand, are delayed by $0.10 \pm 0.02$ (Cowley \&
Crampton  1975)  and $0.08  \pm  0.02$ (LaSala  \& Thorstensen  1985),
indicating that the emission line regions trail the compact object. We
will revisit these emission line velocities in Section 3.3.

Since we have detected  narrow lines from  the mass donor in  emission
rather than absorption, they must originate from the  front half of the
Roche lobe.  The   derived radial velocity  amplitude  is  thus biased
towards  lower values depending   on the detailed  distribution of the
line emission over the Roche  lobe surface.  This  in analogy with the
bias towards larger velocities   when using absorption lines that  are
predominantly formed near the backside of the Roche lobe (e.g. Wade \&
Horne 1988).  The fact that the  radial velocity curve is not strictly
sinusoidal also   suggests   that the   emission    is not  distributed
uniformly.  However, allowing for an  ellipticity in  the fit does not
significantly  improve the $\chi^2$ of the  fit. We can thus constrain
the true velocity  of the centre of mass  of the mass donor ($K_2$) to
be $K_2 > 77.2$ km s$^{-1}$. Without any knowledge of the mass ratio 
and the extent of disc shadowing effects in Sco X-1, it is very difficult
to  quantify the K-correction  that needs  to be  applied to the Bowen
blend velocities  in order to define  the true  $K_2$. A more detailed
modelling of the phase dependence of the  emission from the mass donor,
both in terms of velocity as well as  strength is required, which will
be discussed in a future paper.

\subsection{The radial velocity of the primary}

Direct inspection of the trailed  spectra (Figure \ref{fig2})  clearly
shows  that  the  HeII $\lambda$4686  line   is  anti-phased  with the
CIII/NIII narrow emissions and, therefore, it must  come from near the
compact object.   Significant phase shifts between  the line  core and
centroid have been reported, due  to possible contamination by complex
variable hot-spots (e.g. Cowley \& Crampton 1975).  Therefore, we have
followed the   double-Gaussian  technique  to  estimate  the  velocity
semi-amplitude of  the compact object  $K_1$ from the  line wings (see
Schneider \& Young 1980   for details).  We  employed a  two  Gaussian
band-pass  with  FWHM=100     km s$^{-1}$   and Gaussian    separations
$a=200-1000$ km s$^{-1}$ in steps of 100  km s$^{-1}$. Radial velocity
curves    of different line  sections  yield   consistent results with
velocity   semi   amplitudes in  the   range   50-56 km  s$^{-1}$  and
blue-to-red  crossing phases  0.64-0.60.   Phase delays  of  $0.1-0.2$
relative  to the expected motion  of the compact object, are typically
observed in cataclysmic variables   (CVs),  probably due to   residual
hot-spot contamination.
We have  also attempted  to extrapolate  the  velocity points  towards
phase 0.5 using the light-centre technique (see Marsh 1988 for details)
but the method  fails given the  very small derived  phase drift as we
move from the line core  into the wings.
Applying the  same technique to  the   H$\beta$ line produces  similar
values  for   the line  core,   however   the  phase offset  increases
dramatically in the  line wings. We therefore choose  to  use the HeII
lines as our basis for an estimation of $K_1$.
The bottom  panel of Figure \ref{fig3} presents   the radial velocity  curve of
HeII $\lambda$4686 for $a=600$   km s$^{-1}$, before velocity   points
start to  be corrupted   by continuum noise.  A  formal  sine-wave fit
provides $K=53 \pm 1$ km  s$^{-1}$ which we take as  an upper limit to
the true $K_1$.

\subsection{Doppler mapping}

Many of the narrow lines  that can be identified  in Figure 1, move in
phase with  the secondary star   and  thus are   also  the result   of
reprocessed emission from the irradiated front side of the Roche lobe.
Unfortunately, due to  weakness of the lines, a  Gaussian fit  to each
spectrum is not possible.   Instead, we use Doppler  tomography (Marsh
\& Horne 1988)  to measure the radial velocity  amplitudes of the many
weak  narrow emission lines that are  present. By mapping the observed
data onto a velocity coordinate  frame, Doppler  mapping makes use  of
all data at once, and can thus be used for features  that are too weak
to be separated in each individual spectrum. 
One  effectively resolves  the  distribution of line  emission  in the
corotating frame of the binary system, providing an excellent tool for
identifying the origin and kinematics of the various emission components.
It also allows us to map the emission line distribution from the broad
HeI,HeII  and Balmer  lines.   Secondary   star  emission is   readily
identified in Doppler tomograms  since the solid  body rotation of the
Roche lobe is  mapped  to a  corresponding Roche  lobe area  along the
positive $V_y$ axis.
We   employed a maximum  entropy implementation  of Doppler tomography
whereby  the data is fitted under  the added  constraint of maximising
the entropy,  i.e.  image smoothness  of the  tomogram (Marsh \& Horne
1988). This reduces the  presence of noise  artifacts in the recovered
tomograms and allows a simultaneous fit  to a number of heavily blended
lines.

To check  that the  method  of measuring  the  velocity of  the narrow
emission   lines  using tomography   is  consistent  with   the method
of employing (multiple) Gaussian fits,  we started by producing tomograms
for the  Bowen blend  components.   The Bowen  blend data  was  fitted
simultaneously  with  three Doppler images  corresponding  to the rest
wavelengths of the   three strongest  Bowen  emission components.  The
tomograms revealed the secondary  star emission as  a sharp spot along
the $V_y$ axis, and a  2D Gaussian fit was  used to accurately measure
the position of  the    spot.  This  lead to  inferred   velocities of
$76\pm1$, $76\pm1$  and $77\pm1$  km s$^{-1}$ respectively, confirming
that this method can be used to extract reliable radial velocities.
We then repeated the same procedure for a range of other weak emission
lines present in the spectrum of Sco X-1.  The inferred velocities for
emission from the mass donor spanned from 66  km s$^{-1}$ for SiIII at
4568\AA, 78 km s$^{-1}$ for OII doublet at  4415\AA~to 87 km s$^{-1}$ for
NIII at 4379\AA.  An  example tomogram showing secondary star emission
from the  NIII line  at 4514\AA~is  plotted  in  Figure 4.  The fitted
velocity   based on this  tomogram gives   79 km s$^{-1}$. Unfortunately, the
width of the  narrow components are of  the order of  our instrumental
resolution  and so  are not well  resolved.   However, since only  the
front half of  the Roche lobe  can contribute to the line emission, we can
confidently increase our lower limit to $K_2  > 87$ km s$^{-1}$.  This
technique in conjunction with higher spectral resolution data could be
used to map the emission on the Roche lobe as a function of ionisation
potential such  as done in for example  Harlaftis (1999),  and provide
even more stringent constraints to the true value of $K_2$.


In Figure 4 we also present Doppler tomograms  of the strong and broad
H$\beta$, HeII4686  and HeI4921 lines   for comparison. Clearly  their
line  dynamics   are  radically   different    and  are  dominated  by
contributions from  the  accretion disc  around  the primary. However,
significant asymmetries are present in all lines, leading to localised
hot-spots in  the tomograms. In the case  of  HeII at 4686\AA,  we can
identify  a  clear  emission  contribution  from the  secondary  star,
whereas the Balmer and HeI lines do not show a secondary star emission
component. If anything, the H$\beta$  tomogram indicates a  depression
of line flux at the location of the mass donor. 
The tomograms   directly  illustrate that the   use   of emission line
velocities as   a measure of   the  radial velocity  amplitude of  the
compact object is severely affected and distorted by asymmetries in the
line    profiles.  Indeed,   our   analysis  in  Section 3.3  revealed
significant  phase offsets between  the phasing  of the secondary star
components  and the broad emission  lines. The different phase offsets
that exist  between the velocities of the  H$\beta$ line a versus HeII
are due to the fact that the HeII is mainly distorted by emission from
the mass donor, whereas  the H$\beta$ line is  distorted by a spot  in
the lower left quadrant  of the Doppler image.  The fact that our data
reveals   the same phase   shifts as found  by   Lasala \& Thorstensen
(1985), suggest this asymmetry is a persistent feature in Sco X-1.

The complex  morphology   of the  Doppler  tomograms complicates   the
reliable estimation  of $K_1$.  If the  emitting gas was symmetrically
distributed   around the primary,  the Doppler   image  would reveal a
circular  distribution centred  on   $V_x=0, V_y=-K_1$.   In order  to
estimate   the true  centre   of  the  emission line  distribution, we
subtracted the symmetrical   component from the H$\beta$  Doppler maps
and inspected the residuals. We cycled  the assumed centre of symmetry
between -100 and +25   km  s$^{-1}$ in  steps   of 2 km s$^{-1}$   and
measured the mean  of the residuals  at  high velocities in the  upper
right quadrant which  is not affected by  the spot. The mean residuals
show a pronounced minimum around an assumed centre of symmetry of $V_y
\sim -40$ km s$^{-1}$. This suggests  that the true  value of $K_1$ is
indeed closer to 40 km s$^{-1}$ for Sco X-1.

\section{Discussion}

We have presented the first direct detection  of the mass donor in Sco
X-1  by resolving emission  components originating from the irradiated
front  side of  its  Roche lobe.   Radial velocity  fits  were used to
establish    an absolute orbital    ephemeris and measure the systemic
velocity to  high accuracy; $\gamma = -113.8  \pm 0.6$ km s$^{-1}$.  A lower
limit to the radial velocity of the mass donor is given by the radial
velocity amplitude of the  sharp Bowen blend emission components which
leads to $K_2 > 77$ km s$^{-1}$. Detection of secondary star emission at even
higher  velocities in other transitions lead  to a more stringent lower
limit of $K_2 > 87$ km s$^{-1}$.
The   strong broad emission  line  show   orbital  motions roughly  in
anti-phase  with the  secondary  star.  We   confirm significant phase
shifts between the radial velocity curves  of the strong lines and the
absolute  ephemeris.   Doppler  tomography reveals  that these  phase
shifts  are  due  to  persistent asymmetries   in  the  emission  line
distribution around the  compact object.   The formal double  Gaussian
method leads to  a  firm upper limit to   the primary radial  velocity
amplitude  of $K_1 <  53$  km s$^{-1}$. A more  reliable value  is derived by
searching  for the optimal  centre  of symmetry  in the accretion  disc
emission which leads to an estimated  $K_1$ value of $40 \pm 5$ km s$^{-1}$.
Our absolute ephemeris confirms previous suggestions that the phase of
optical minimum light corresponds closely to  orbital phase zero, when
the contribution from the heated mass donor is smallest.

The combination  of our upper limit  to $K_1$ and at  the same  time a
lower limit to $K_2$ leads to  a firm limit to the  mass ratio of $q =
M_2/M_1 <  0.69$ when using  the formal values  of $K_1 < 53 $ and  $K_2 > 77$ as
derived from the Gaussian fits. Using the more restrictive values $K_1
\sim 40$, $K_2 > 87$, we derive a limit of $q \la 0.46$.
%
Another constraint is  given by  the fact  that the observed  emission
originates from the front half of the Roche lobe and its velocity thus
lies in between that of the  true $K_2$ value and  the velocity of the
L1 point ($V_{L1}$). Since  we observe emission at velocities as
low as 66 km s$^{-1}$, we can thus derive an upper limit on $K_2$ for a given
mass ratio  in order to   make sure that  the velocity  of L1  is  not
greater than 66 km s$^{-1}$.  In Figure 5 we plot the various limits on $K_2$
as a  function of  mass   ratio.  As can be   seen  in the plot,   the
combination of  $V_{L1}<66$ and $K_2<K_1/q$ leads to  $K_2 < 148$ km s$^{-1}$
regardless of the mass ratio.
This upper limit proves   interesting from the  point  of view of  the
recent inclination estimate based on the active radio lobes of Sco X-1
(Fomalont  et al.   2001).  If   the orientation of   the radio  lobes
corresponds to the rotation axis of the orbital plane, we can estimate
the component masses using Kepler's  law.  With a typical neutron star
mass of $M_1 \sim 1.4  M_{\sun}$, this inclination ($44\pm6^{\circ}$)
predicts  relatively high $K_2$ values across   the allowed mass ratio
range (Figure \ref{fig5}).
This suggests  that a slightly  lower  orbital inclination  seems more
likely  unless  the neutron star is    atypical and significantly less
massive than the Chandresekhar limit.   However, given the uncertainty
on  the inclination, all  constraints are still  compatible with a 1.4
$M_{\sun}$ neutron star within 1$\sigma$.
A likely set  of parameters for Sco X-1  given the various constraints
could   be  $K_1=40$ km   s$^{-1}$,   $K_2=133$ km  s$^{-1}$,   $q=0.30$
corresponding to  component  masses of  $M_1=1.4M_{\sun}$  , $M_2=0.42
M_{\sun}$ for an orbital inclination of $38^{\circ}$. 
Given the orbital  period of Sco X-1,  the companion star must then be
a significantly evolved subgiant  having a radius of  more than twice of
the   corresponding main sequence radius of   $\sim 0.5R_{\sun}$.  Its
mean density is solely determined   by the orbital period which  gives
$\rho=0.31 $g  cm$^{-3}$ compared  to  $\sim  5.6$  g cm$^{-3}$ for  a
$0.42M_{\sun}$ main sequence star.
We note that these parameters are also consistent with Sco X-1 being a
persistent X-ray  binary, with  an  approximate mass  transfer rate of
$\dot{M}_2   \sim   8 \times   10^{-10} M_{\sun}    yr^{-1}$ using the
prescription from  King, Kolb \&  Burberi (1996).   This is  enough to
maintain the  irradiated-disc hotter than  ~6500 K everywhere avoiding
global instabilities and thus outbursts.

Bandyopadhyay et al.  (1997,1999) fail to detect any features from the
companion  star in their IR  spectra of Sco  X-1 at  two epochs.  Most
notably, the CO band-head  near 2.2$\micron$ that  is present in  late
type stars is absent.  However, the appearance of  the donor star will
be affected by  irradiation and  both  epochs were  taken near orbital
phase 0.5  (using our derived ephemeris) when  these  effects would be
particulary severe.  In addition, the use of the CO-bands for spectral
type  classification can  be misleading   because   their strength  is
strongly dependent  on effective temperature and, furthermore,   the  C,O
abundances can be   significantly  non-solar for   companion  stars in
interactive binaries (e.g.  Howell 2001).
In order to derive more accurate system parameters, we need to be able
to model the K-correction, and use higher  spectral resolution data to
resolve the emission across the Roche lobe  for a range of transitions
using the  methods  described in this   paper.  Detecting the companion
star in  absorption should be  possible around true orbital phase zero
when the effects of X-ray heating are minimal.

\acknowledgments

DS is  supported by a PPARC Fellowship.  Use of  the MOLLY and DOPPLER
software  developed by T. R.   Marsh  is gratefully acknowledged.  The
William Hershel Telescope is operated on the island of La Palma by the
Isaac Newton Group  in  the Spanish  Observatorio   del Roque  de  los
Muchachos of the Instituto de Astrofisica de Canarias.


\clearpage

\begin{table*}
\begin{center}
\caption{Journal of observations}
\vspace{.1cm}
\begin{tabular}{ccccc}
\tableline\tableline
Date & UT interval & Number of spectra & Exposure time & Orbital phase interval \\
\tableline
28/6/1999	&	21:55 -- 02:10	&	40 & 300s & 0.81-1.03\\
29/6/1999	&	20:55 -- 02:15  &	51 & 300s & 0.03-0.31\\
30/6/1999	&	20:56 -- 01:55	&	46 & 300s & 0.30-0.56\\
\tableline
\end{tabular}
\end{center}
\end{table*}

\clearpage

\begin{figure}
\psfig{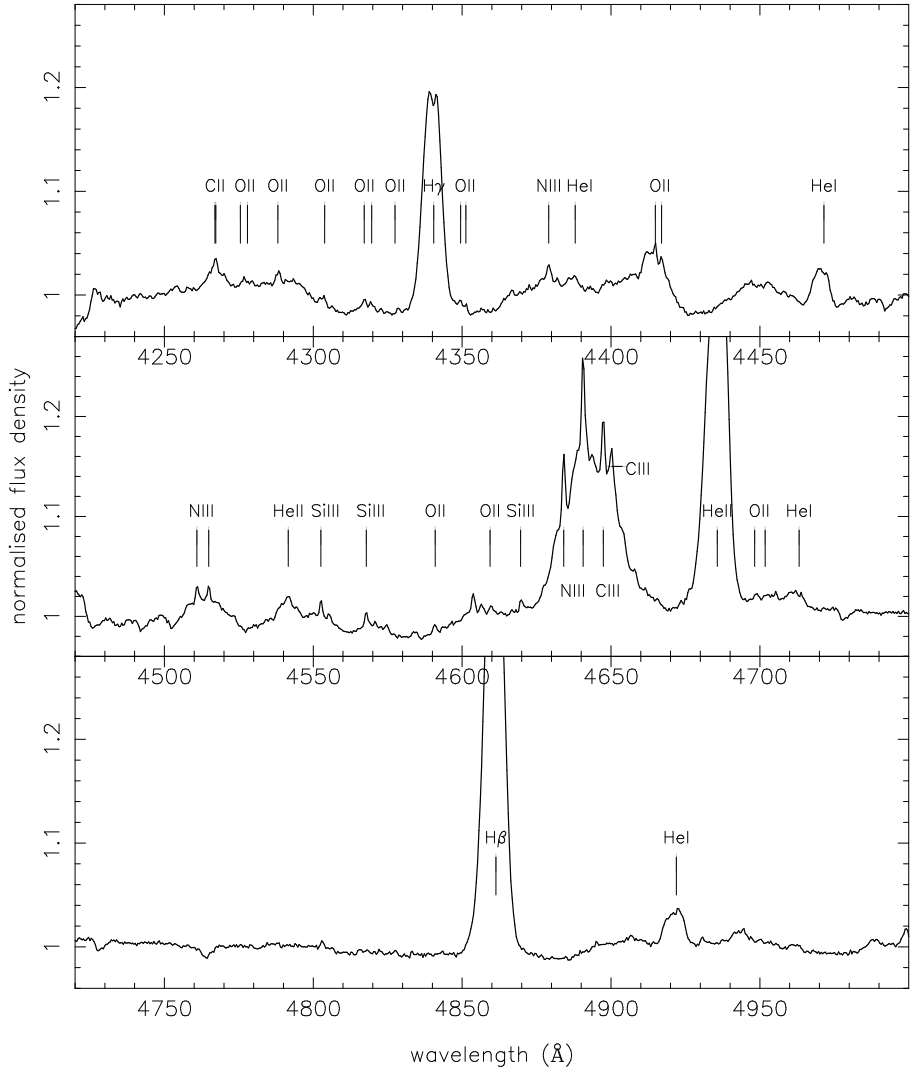}
\caption{
The average normalised spectrum of  Sco X-1 in  the rest frame of  the
donor star. The three panels are on the  same intensity scale with the
strong HeII4686 and H$\beta$ going  off-scale in order to highlight the
weaker lines. The peak strength in HeII is around 1.46 compared to the
continuum, whereas it is 1.39 for H$\beta$. \label{fig1}}
\end{figure}

\clearpage

\begin{figure}
\psfig{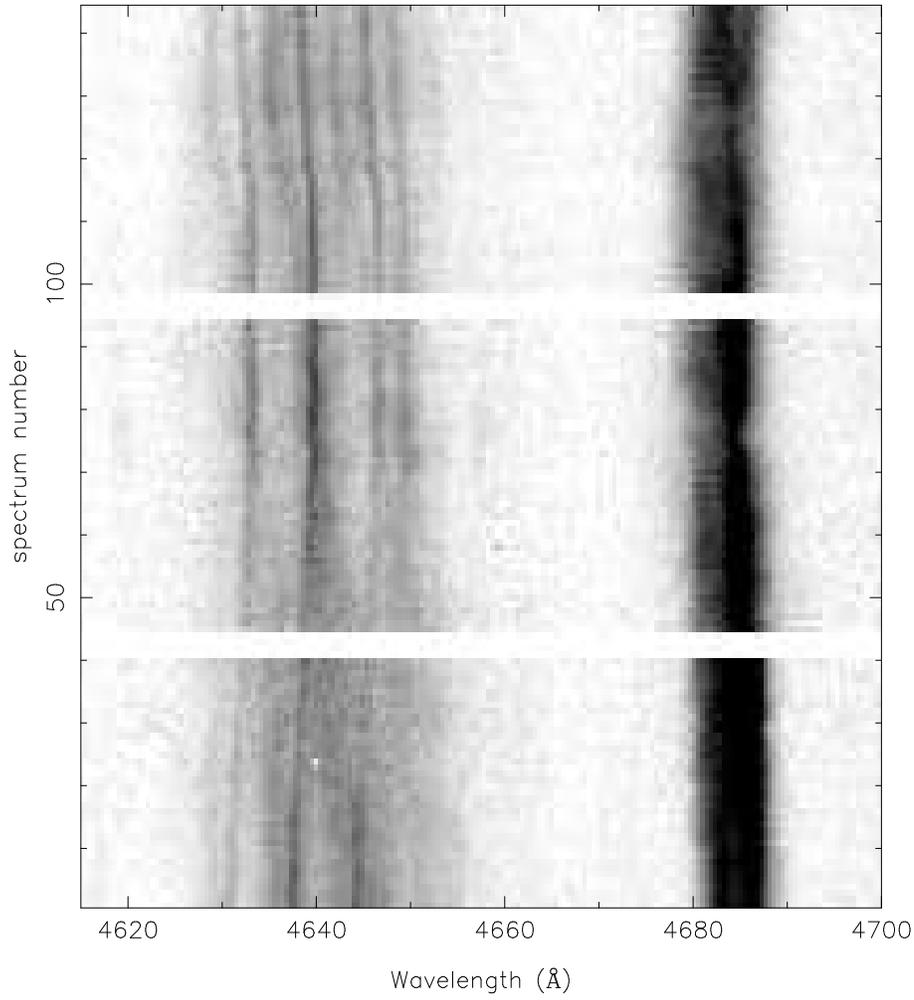}
\caption{Trailed spectrogram of the Bowen blend and HeII4686. Three sharp components moving in phase with each other stand out from the broad Bowen blend emission. Small gaps separate the data from the three nights.   \label{fig2}}
\end{figure}

\clearpage
\begin{figure}
\psfig{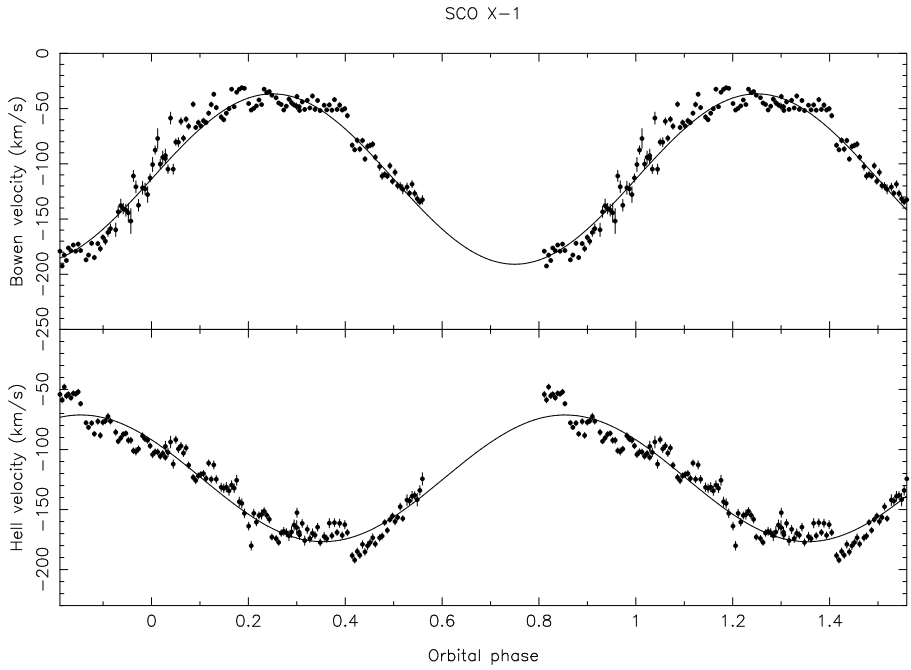}
\caption{
Top:   radial  velocity curve   of the  narrow   Bowen blend  emission
components as derived  from a multiple  Gaussian fit. Data is repeated
over   two cycles. Best   sinusoidal fit  is   over plotted as  a solid
line. The orbital phase of the binary is calculated using the best fit
radial  velocity  curve to determine the    zero point. Bottom: Radial
velocity  curve of the HeII emission  line based  on a double Gaussian
fit with separation 600  km s$^{-1}$ and FWHM of 100  km s$^{-1}$. Sinusoidal fit is
over plotted indicating a phase shift of 0.6 with  respect to the donor
star emission.  \label{fig3}}
\end{figure}

\clearpage
\begin{figure}
\psfig{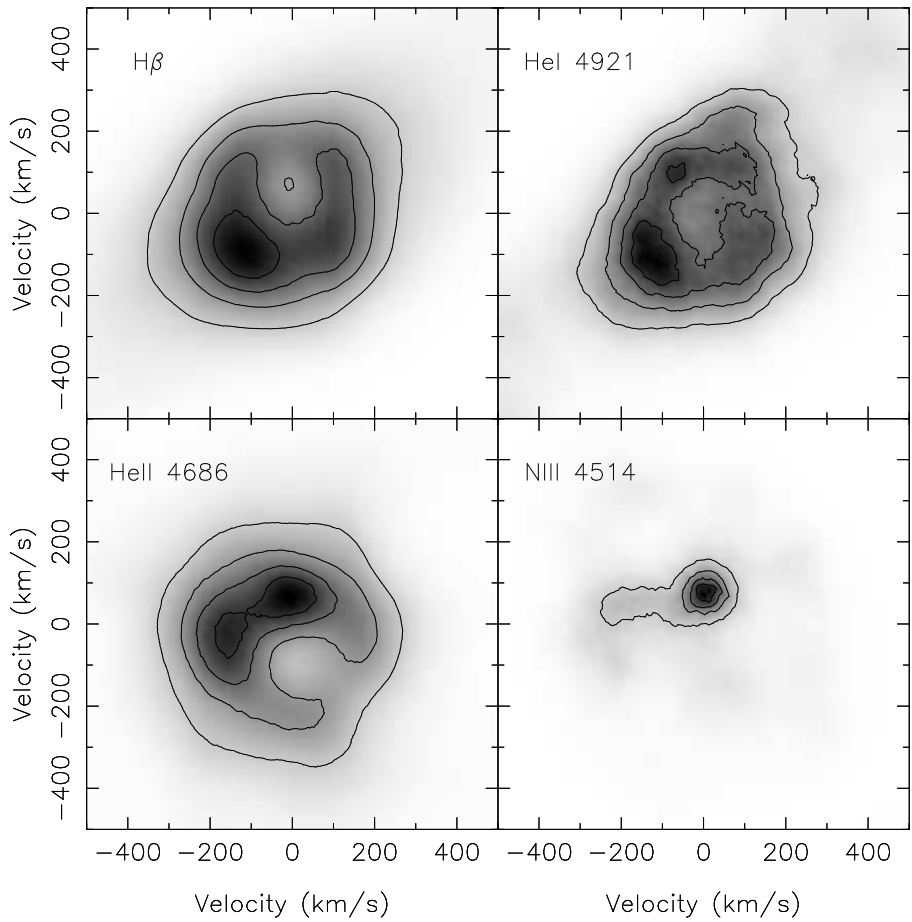}
\caption{Doppler tomograms reconstructed from the H$\beta$, HeI4921, HeII4686 and NIII 4514 lines using maximum entropy optimisation. Both grayscale as well as contours span the full scale of the maps. The origin corresponds to the centre of mass of the binary system using $\gamma=-113.8$ km s$^{-1}$.  \label{fig4}}
\end{figure}

\clearpage

\begin{figure}
\psfig{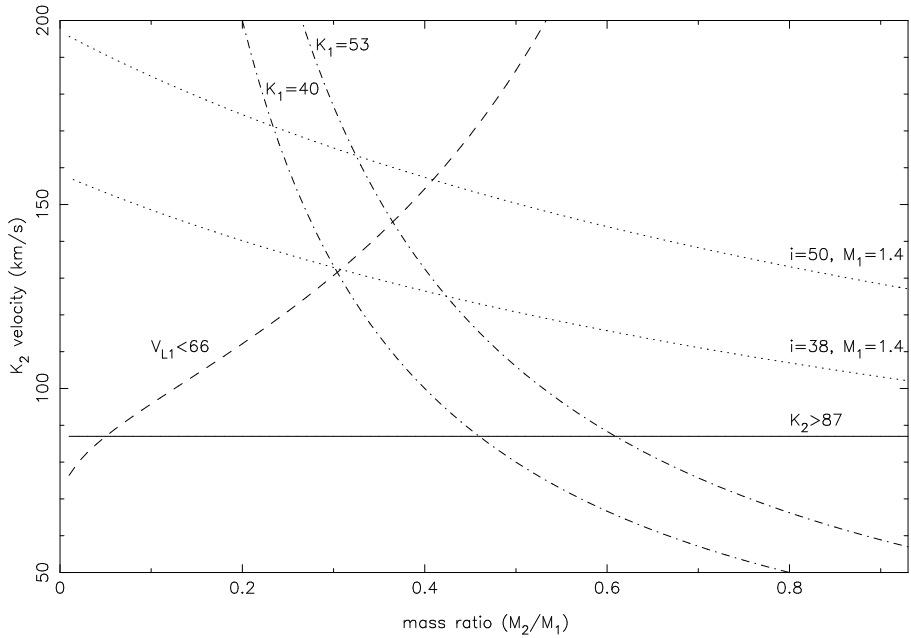}
\caption{
Plotted are constraints to  the value of $K_2$ as  a function of  mass
ratio.  The horizontal solid line is the lower limit to $K_2$ as
derived from the observed line emission ($K_2 > 87$). The  dot-dashed lines denote  the
upper limits on  $K_2$ given  that $K_2  < K_1/q$  for $K_1=53$ (upper
curve) and $K_1$=40 km s$^{-1}$.  The requirement that the  velocity of the L1
point may not exceed 66 km s$^{-1}$ leads to the dashed curve. The two dotted curves
denote the   values of $K_2$  if  the primary is  a 1.4  $M_{\sun}$ neutron
star. The two   curves correspond  to  $\pm$1$\sigma$ limits for  the orbital inclination based on the orientation of the radio lobes ($44\pm6^{\circ}$).}
\label{fig5}
\end{figure}




\begin{thebibliography}{}

\bibitem[]{}Augusteijn, T. et al. 1992, A\&A, 265, 177
\bibitem[]{}Bandyopadhyay, R.M., Shahbaz, T., Charles, P.A., Naylor, T., 1999, MNRAS, 306, 417
\bibitem[]{}Bandyopadhyay, R.M., Shahbaz, T., Charles, P. A., van Kerkwijk, M. H., Naylor, T., 1997, MNRAS, 285, 718
\bibitem[]{}Bradshaw, C.F., Fomalont, E.B., Geldzahler, B.J., 1999, ApJ, 512, L121 
\bibitem[]{}Cowley, A.P, Crampton, D., 1975, ApJ, 201, L65
\bibitem[]{}Fomalont, E.B., Geldzahler, B.J. Bradshaw, C.F., 2001, ApJ, in press 
\bibitem[]{}Gottlieb, E.W., Wright, E.L., Liller, W., 1975, ApJ, 195, L33
\bibitem[]{}Harlaftis, E.T. 1999, A\&A, 346, 73
\bibitem[]{}Harlaftis, E. T., Steeghs, D., Horne, K., Martin, E. \& Magazzu, A.
 1999, MNRAS, 306, 348
\bibitem[]{}Howell, S., 2001, PASJ, in press (astro-ph/0106214)
\bibitem[]{}Kaler, J.B., 1976, ApJSS, 31,517
\bibitem[]{}King, A. R, Kolb, U., Burderi, L., 1996, ApJ, 464, L127
\bibitem[]{}LaSala, J., Thorstensen, J.R., 1985, AJ, 90, 207 
\bibitem[]{}Marsh, T.R., Horne, K., 1988, MNRAS, 235, 269 
\bibitem[]{}Marsh, T.R. 1988, MNRAS, 231, 1117  
\bibitem[]{}Oke, J.B. 1990, AJ, 99, 1621
\bibitem[]{}Sandage, A.R. et al. 1966, ApJ, 146, 316 
\bibitem[]{}Schachter, J., Filippenko, A.V., Kahn, S.M., 1989, ApJ, 340, 1049
\bibitem[]{}Schneider, D.P., Young, P., 1980, ApJ, 238, 946 
\bibitem[]{}Schwope, A.D., Catalan, M.S., Beuermann, K., Metzner, A., Smith, R.C. \& Steeghs, D. 2000, MNRAS, 313, 533
\bibitem[]{}Still, M.D., Quaintrell, H., Roche, P. D. \& Reynolds, A. P. 1997, MNRAS, 292, 52
\bibitem[]{}Tinbergen, J. \& Rutten, R. 1992, La Palma ING User Manual 21, Isaac Newton Group, La Palma
\bibitem[]{}Wade, R.A. \& Horne, K. 1988, ApJ, 324, 411
\bibitem[]{}Westphal, J.A., Sandage, A. \& Kristian, J. 1968, ApJ, 154, 139 
\bibitem[]{}Willis, A.J. et al. 1980, ApJ, 237, 596

\end{thebibliography}
\end{document}